\documentclass[12pt]{iopart}
\usepackage{graphicx}
\usepackage{amssymb}
\usepackage{cite}
\begin{document}

\title[Dust as plasma probe]{Dust as probe for horizontal field distribution in low pressure gas discharges}

\author{Peter Hartmann$^{1,2}$, Anik\'o Zs. Kov\'acs$^1$, Jorge C. Reyes$^2$, Lorin~S.~Matthews$^2$, and Truell W. Hyde$^2$}

\address{                    
 $^1$ Institute for Solid State Physics and Optics, Wigner Research Centre, Hungarian Academy of Sciences, P.O.B. 49, H-1525 Budapest, Hungary\\
 $^2$ Center for Astrophysics, Space Physics and Engineering Research (CASPER), One Bear Place 97310, Baylor University, Waco, TX 76798, USA
}
  
\begin{abstract}
Using dust grains as probes in gas discharge plasma is a very promising, but at the same time very challenging method, as the individual external control of dust grains has to be solved. We propose and demonstrate the applicability of the RotoDust experiment, where the well controlled centrifugal force is balanced by the horizontal confinement field in plane electrode argon radio frequency gas discharges. We have reached a resolution of 0.1~V/cm for the electric field. This technique is used to verify numerical simulations and to map symmetry properties of the confinement in dusty plasma experiments using a glass box.

\end{abstract}

\maketitle

\section{Introduction}

Since the discovery and standard use of dusty plasmas, one of the most promising applications is the use of micron-sized solid particles as discharge probes. This concept is based on the assumption that the particles will settle at positions where the forces acting on them are in balance, and that a low enough dust number density (down to a single particle) will not alter the discharge conditions. At the same time, the dust particles are big enough to be easily observed by conventional video microscopy. After the formulation of the concept \cite{Allen00,Samarian05}, much effort has gone into the elaboration of the details of this technique in recent years \cite{Basner09,Maurer11,Beckers,Carstensen11,Schubert12}, but it is still an exotic method with large, but untapped potential. For this technique to become widely useful, the issue of controlling the forces acting on the particles has to be solved. 

A single dust particle immersed and charged in a low pressure gas discharge is exposed to a series of interactions, which include:
\begin{itemize}
\item gravity, $F_G=mg$,
\item electric field, $F_E=qE$,
\item magnetic field, $F_M=q({\bf v}\times{\bf B})$,
\item ion drag, $F_I \approx Z z n_i \alpha {\bf u}$,
\item neutral drag $F_N \approx -m \nu {\bf w}$,
\item thermophoresis, $F_T \approx 2.4 n_g \pi \lambda_g (\partial T_g/\partial{\bf r})$, etc.,
\end{itemize} 
where $m$ is the mass, $q=-eZ$ is the charge, and ${\bf v}$ is the velocity of the dust grain, $z=Ze^2/(4\pi\varepsilon_0 a T_e)$ is the dimensionless grain charge, $T_e$ is the electron temperature, $a$ is the grain diameter, $n_i$ is the ion density, $\alpha$ is a phenomenological drag coefficient, ${\bf u}$ is the velocity of the ions relative to the grain, $\nu$ is the Epstein drag coefficient, ${\bf w}$ is the grain velocity relative to the background gas, $n_g$ is the density of the background gas, $\lambda_g$ is the collisional mean free path of the background gas atoms and $T_g$ is the gas temperature. Extensive discussions of all these interactions and the dust particle charging can be found in \cite{Beckers,Vadim}.

The order of importance of the interactions listed above strongly depends on the local discharge conditions at the position of the dust particle. For instance, in the sheath region of a radio frequency (RF) noble gas discharge, the levitation height is mostly determined by the balance of gravity and the strong sheath electric field, while in the bulk of an RF discharge under microgravity conditions, the observed dust void forms due to the strong ion drag effect \cite{Melzer07}. 

The limitations of using a dust grain as a plasma probe originate from the fact that once it is released into the plasma, we lose control over its properties, like electric charge (surface potential) and position in the plasma. An important capability is to change our possibilities from simply observing the particle to being able to manipulate it in a controlled manner. Laser manipulation, being an extremely successful technique in other branches of dusty plasma research \cite{Stoffels,Melzer01,Schneider13}, provides a possibility to push individual dust grains; however, the force introduced by the radiation pressure of a laser beam on a grain can only be estimated (rather inaccurately) and poorly controlled over an extended period of time. 

A different, more promising direction seemed to be to modify the gravitational force $F_G$. An obvious way of realizing this is to change the mass of the particle by simply choosing different sized dust grains. Practical reasons limit the dust size in the range of $1<a<50~\mu$m. Smaller particles are undetectable, while bigger particles are to heavy to be levitated by the fields in the discharge. Changing the physical conditions of the grain e.g. from solid to hollow may extend the size range a bit. More innovative approaches target the gravitational acceleration constant $g$. Micro-gravity experiments on parabolic flights and the International Space Station (ISS) can practically reach $g\approx0$~m/s$^2$ \cite{Nefedov,PK3P}. Alternatively, hyper-gravity experiments in centrifuges could increase the gravitational acceleration up to $g\approx26.5$~m/s$^2$ \cite{Beckers11}. These methods make it possible to move the dust across the discharge from the central, practically electric field free region to deep into the high field sheath (edge) plasma in a controlled way. Both methods share the advantage that the discharge plasma is largely unaffected while the dust grain is moved, however, they rely on very expensive and widely inaccessible infrastructure, making each experiment unique and practically irreproducible.

Here we propose a new method to measure the horizontal electric field in typical RF low pressure gas discharges with electrode configurations symmetric around a vertical axes. In section 2 the experimental method and setup are described in detail, in section 3 the experimental results are presented and verified against simulation results, and an example for application is given as well, followed by a short summary in section 4.

\section{Experimental method}

Our method of introducing a well controlled external force to the dust grain in the discharge plasma utilizes the centripetal force, which keeps a particle gyrating around a fixed center on a circular orbit. The rotation is introduced externally, while the centripetal force is provided by the horizontal component of the electric field inside the discharge.

\begin{figure}[htb]
\centering
\includegraphics[width=0.8 \columnwidth]{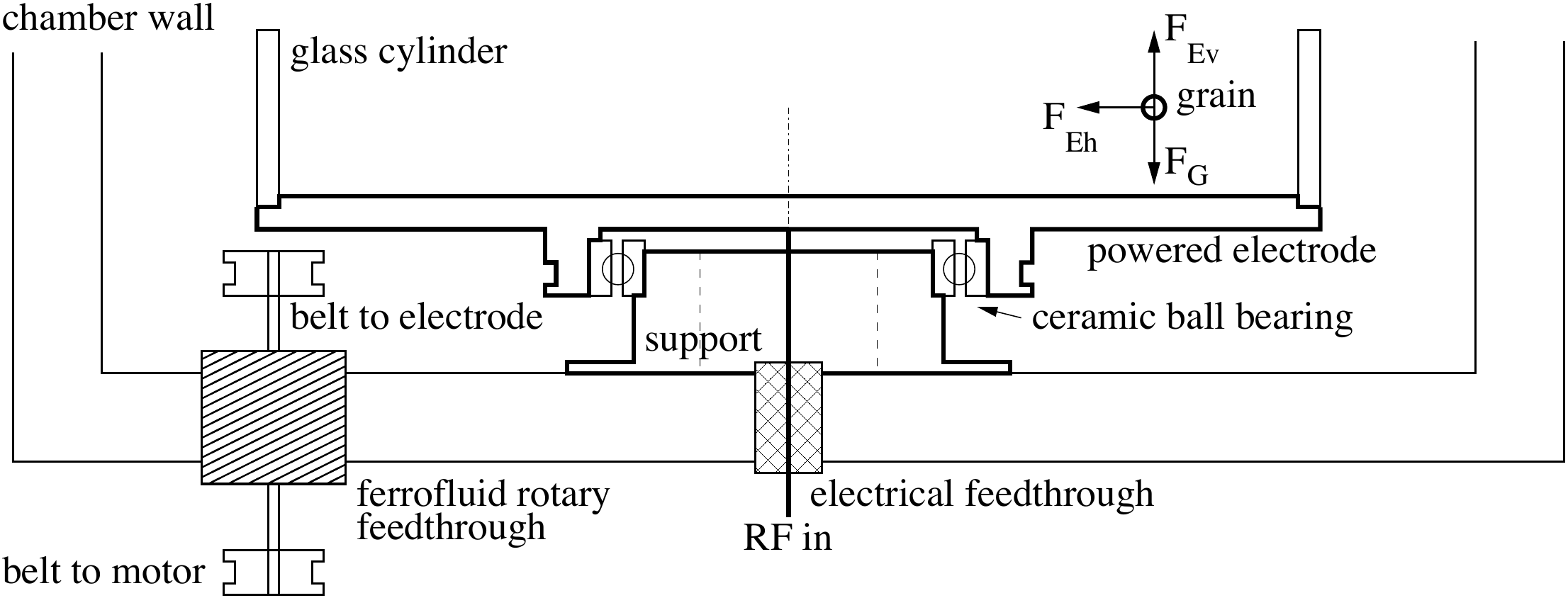}
\caption{\label{fig:setup} 
Schematics of the ``RotoDust'' electrode configuration. The main forces of interest: gravity, horizontal and vertical components of the electric field are illustrated acting on a representative dust grain.} 
\end{figure}

The experiments were performed in our dusty plasma setup already introduced in \cite{Harti11}. The lower electrode has been modified according to figure~\ref{fig:setup}, showing the schematic sketch of our ``RotoDust'' setup \cite{RotoDust}. An adjustable speed dc motor was used to drive the powered electrode through a ferrofluid rotary feedthrough and Viton belts. The angular velocity of the rotating electrode could be continuously adjusted in the range $0\le\Omega<25$~rad/s. The rotating electrode drags the background gas above it and induces rotation of the gas. The rotating gas, on the other hand, drags the dust particles, resulting in a stable rotation of the dust component, which builds up in a few seconds after switching on the rotation. Due to the low pressure in the 1 Pa range used in our experiments, it is safe to assume that turbulence does not develop. Also it seems to be certain that this rotation of the background gas does not affect the plasma properties, because electron and ion transport in these discharges happens on the sub micro-second timescale, and the uncorrelated thermal velocity of the background gas atoms exceeds the rotation speed by about 3 orders of magnitude at room temperature.  

During the measurement, we introduce a single, size-certified spherical melamine-formaldehyde (MF) dust particle into the discharge, switch on rotation at a given, constant angular velocity $\Omega$ and measure the radius $R$ of the stable circular orbit of the dust particle. In this case, observing the motion from the laboratory frame, the centripetal force, which is purely horizontal, is provided by the horizontal component of the confinement electric field $E_{\rm conf}$ at a radial position $R$:
\begin{equation}\label{eq:equ}
qE_{\rm conf}(R)=mR\Omega^2.
\end{equation}
All factors on the right hand side of (\ref{eq:equ}) are known or measured with high precision, thus the product of the electric charge of the grain and the confinement field strength is found with the same high precision. As a consequence, absolute field strength measurements are as accurate as the grain charge estimate is accurate, which is typically in the 10-20~\% range. For relative field measurements, however, the stability of the grain charge is important, while its absolute value together with its uncertainty is irrelevant, resulting in significantly better relative accuracy. To estimate the sensitivity of this method we assume typical values of dust grain mass ($m \approx 10^{-13}$~kg for a grain with 5~$\mu$m diameter) and charge ($Z\approx -10^4e$) based on earlier experiments. Further assuming angular velocities in the range of 1~rad/s and orbit radii in the range of centimeters, the detectable electric fields are in the order of V/m (or $10^{-2}$~V/cm). This sensitivity exceeds that of every conventional diagnostic technique (e.g. electric probes, Stark spectroscopy, etc.), by orders of magnitude. This enhanced sensitivity makes this technique attractive for many applications and drives recent developments, including the present work.

\section{Results and application}

In the first experiment, which targets the demonstration and validation of the method, we have used MF dust grains with $a=9.16~\mu$m diameter ($m=6.03\times10^{-13}$~kg), which were inserted into a 13.56 MHz, radio frequency (RF), $p=1.0$~Pa pressure argon gas discharge, absorbing 8 Watts of RF~power with peak-to-peak voltage of $V_{\rm pp}=175$~V. The lower, flat, horizontal, powered electrode had a glass ring of $R_{\rm wall}=85$~mm inner radius and $H=100$~mm height at its edge. The top plate of the grounded vacuum chamber was 120~mm above the powered electrode.

\begin{figure}[htb]
\centering
\includegraphics[width=0.7 \columnwidth]{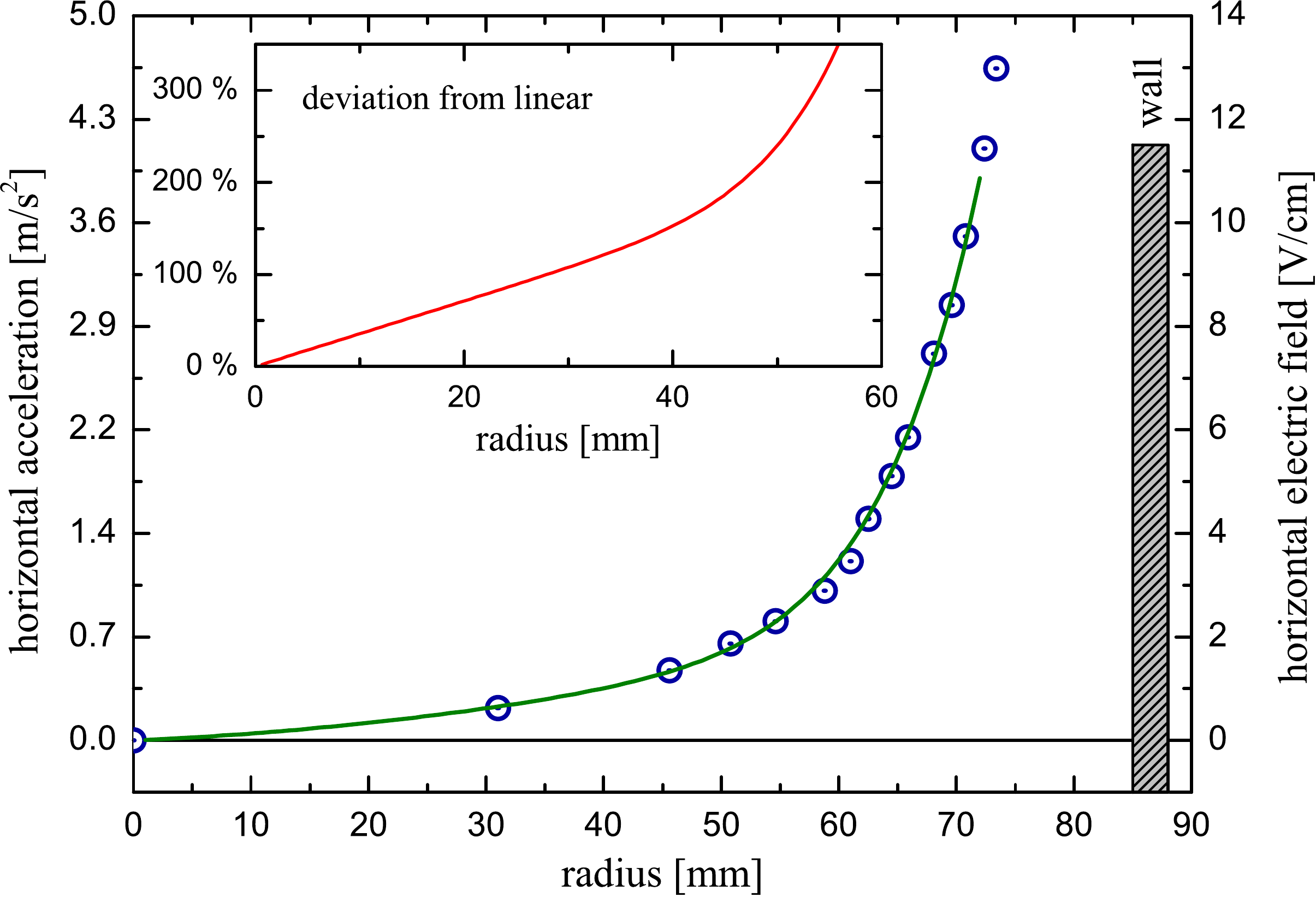}
\caption{\label{fig:9um} 
(Color online) Measured radial distribution of the horizontal centripetal acceleration (left scale) and electric field (right scale). Lines represent polynomial fits, see text for details. The inset shows the ratio of higher order and the first order terms of the fitted polynomial.} 
\end{figure}

During the actual measurements a few ($<10$) particles were dropped into the discharge. After setting a stable rotation speed, the orbital radii were measured with the aid of a laser source mounted on a linear translator stage. The raw measured quantities are the angular velocities $\Omega$ and radii of the orbiting particles $R$. From these the centripetal acceleration $g_{\rm centr}=R\Omega^2$ and further the horizontal confinement electric fields are derived by applying eq. (\ref{eq:equ}) (see fig.~\ref{fig:9um}). 

As expected, at rest ($\Omega=0$) the equilibrium position of our dust grains is at the center. Slowly increasing the rotation rate, the grains are observe to react very sensitively, indicating a very weak confinement force near the center. To approach the glass wall, which is negatively charged, much higher rotation rates are needed. In this case we have reached rotation rates up to 1.8~rot/s ($\Omega=11.3$~rad/s). In fig.~\ref{fig:9um} a polynomial fit in the form $E_{\rm conf}(R)\approx c_1R+c_2R^2+c_9R^9$ is also shown, which is used for further analysis. First we test the assumption, often made while discussing trapped systems, that the trap can be modeled by a harmonic potential close to the center. The inset in fig.~\ref{fig:9um} shows the ratio $(c_2R^2+c_9R^9)/(c_1R)$, quantifying the deviation of the measured force field from an ideal harmonic trap (which has a linear force field). We see that at $R\approx25$~mm $(R/R_{\rm wall}\approx 0.3)$ the second order contribution becomes equal to the linear term and above $R\approx45$~mm $(R/R_{\rm wall}\approx 0.5)$ the higher order term dominates. 

The most critical unknown parameter linking the centripetal acceleration to the electric field (left and right scales of fig.~\ref{fig:9um}) is the charge of the dust grain. First of all, we assume the charge is constant and independent of the radial coordinate $R$. This approximation will obviously fail very close to the glass wall, but the radial extent for which this assumption is valid will be discussed later. Second, we have to quantify the electric charge of the dust grain. Luckily, over the past two decades of intensive dusty plasma research, several experimental and theoretical methods have been developed to do this. In Ref.~\cite{RotoDust} we describe the application of two experimental methods. In the case of a large plasma crystal, the fluctuation spectra are used to determine the charge, while in the case of a two particle system, the center of mass and inter-particle oscillations are used to extract the charge. Both of these methods rely on particle tracking velocimetry (PTV) \cite{Feng07} and the assumption of screened Coulomb (Yukawa) type inter-particle forces. This approximation is widely accepted and confirmed in the case of (horizontal) single layer configurations, but it is known to fail in the vertical direction due to streaming of charged plasma particles. Based on our experience with our experimental setup, we estimate $Z=q/e=-12000\pm2500$ in the current experiment, but this estimate will be refined later in this manuscript. For alternative methods and discussions see e.g. Ref.~\cite{Sharma12,Carstensen10}. 

From the theoretical side, for the discharge conditions relevant to dusty plasma experiments, the orbital motion limited (OML) analytical method is preferred for calculating the charge over other approaches, like the diffusion limited model (DLM) or the radial drift model (RDL). For detailed discussion and derivation of these approaches, see Ref.~\cite{Vadim}. Besides analytical models, phenomenological approximations and numerical simulations can also be applied to calculate the dust grain charge \cite{Miloch09,Angela11,Miloch12}.     

\subsection{Numerical simulation}

One of the most promising applications of using dust grains as plasma probes is to use experimentally obtained electric field data to validate numerical simulations, where the electric field (through the potential) is usually one of the principle quantities implicitly calculated during the simulations. To demonstrate the power of this technique, we have modified our ``particle-in-cell with Monte Carlo collisions'' (PIC/MCC) code \cite{Donko11}, which has been validated and benchmarked in a multi-group collaboration \cite{Derzsi13}. We have extended it to describe cylindrically symmetric finite discharges, including wall charging and electron--argon collisions \cite{Phelps94,Phelps99}. The description of all details of this simulation code goes well beyond the scope of this paper, here we give only a brief introduction. The simulation results have been calculated using our electrostatic 2d(R,H)3v PIC/MCC code. We restricted the simulation volume to the relevant region inside the confining glass cylinder, which is placed at the electrode edge (see Fig.~\ref{fig:setup}). Input parameters were set based on the experiment: Voltage amplitude = 170~V at 13.56~MHz, Argon pressure = 1~Pa, gas temperature = 400~K. The superparticles of the PIC simulation had a constant weight of 80\,000 for both electrons and Ar$^+$ ions, the computational grid was set to represent constant volume cells and had a size of $512\times512$. The electrostatic Poisson's equation was solved in every time-step using the combined Fourier (in $H$) and direct space (in $R$) method described in \cite{Chao97}. Computational speedup was achieved by implementing MPI parallelization. It is to note, that PIC/MCC simulations have already been successfully used to compute forces acting on dust particles if different geometries, like in \cite{Schweigert08,Schubert11}.

\begin{figure}[htb]
\centering
\includegraphics[width=0.8 \columnwidth]{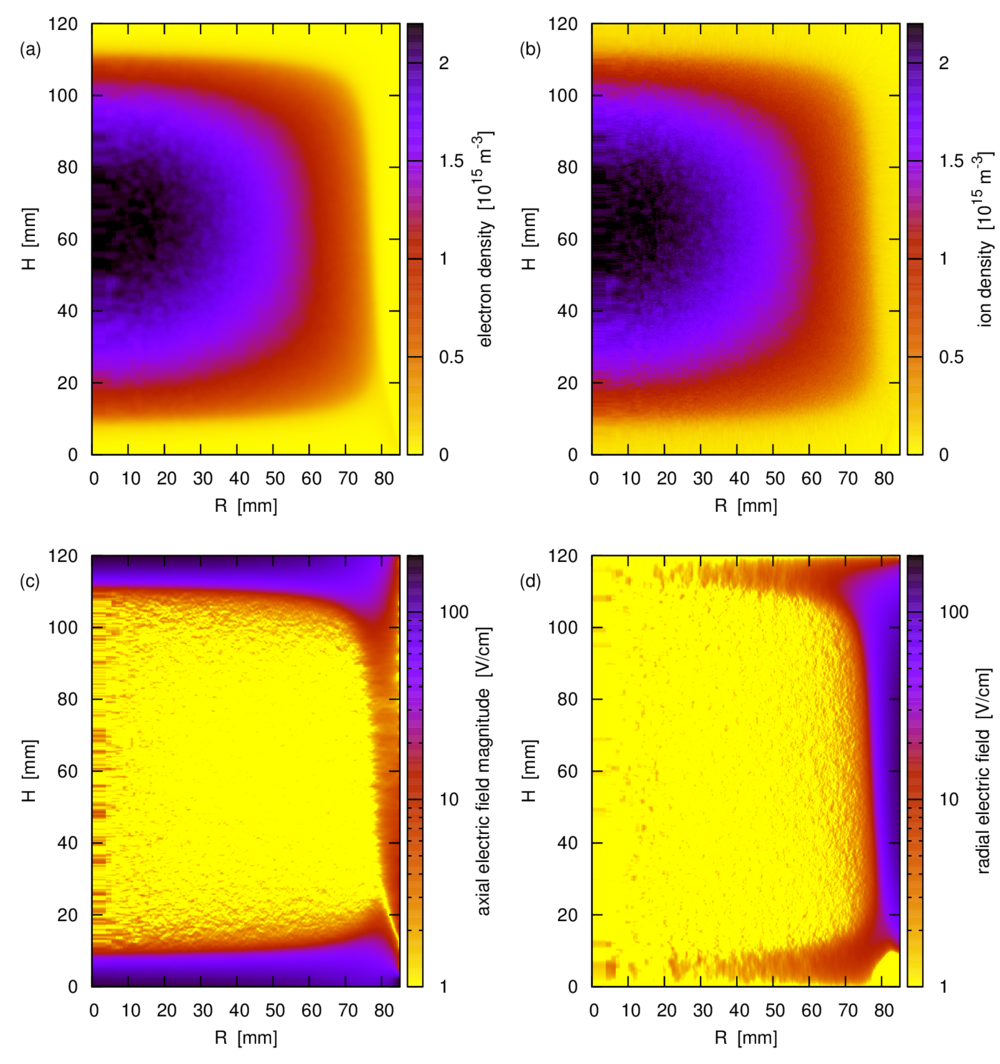}
\caption{\label{fig:pic} 
(Color online) Cylindrical PIC/MCC simulation results: (a) electron density, (b) Ar$^+$ ion density, (c) magnitude (absolute value) of the axial electric field, and (d) radial electric field. Horizontal axis shows the radial distance from the center, while the vertical axis runs from the lower, powered electrode to the grounded top electrode of the discharge. The distributions are averaged over ten RF cycles.} 
\end{figure}

Figure~\ref{fig:pic} shows time averaged distributions of electron and ion densities, and the axial and radial electric fields. The presence of a large density, practically field free bulk plasma is the most prominent feature, clearly visible in the figures. This bulk plasma is surrounded by boundary layers, referred to as plasma sheaths, on all sides. The axial electrode sheath is a well studied feature, accurately described by earlier simulations and analytical models \cite{Schulze08,Schulze11}. Most relevant here are panels (c) and (d), as the axial field is responsible for the levitation of the dust grain through the simple relation 
\begin{equation}\label{eq:ax}
q{\bf E}_{\rm axial}(H)+m{\bf g}=0,
\end{equation}
expressing vertical force balance at vertical position $H$, while the radial field yields the confinement. 

Here we return to two questions posed above related to the dust grain charge. Our first assumption was that the charge is independent of the radial position $R$. From figure~\ref{fig:pic}(a,b) we can see that at the lower sheath edge (around $10<H<15$~mm height) the charged particle densities (relevant for the charging process) are fairly constant up to $R\approx75$~mm, supporting our assumption and the results shown in figure~\ref{fig:9um}.

To refine our first estimate for the dust grain charge, we iteratively converge its value appearing in eqs.~(\ref{eq:equ}) and (\ref{eq:ax}) using the data from our PIC/MCC simulation for the electric field components, and the measured centripetal acceleration data. This procedure results in the levitation height $H(R)$ and the constant charge $q$. 

\begin{figure}[htb]
\centering
\includegraphics[width=0.7 \columnwidth]{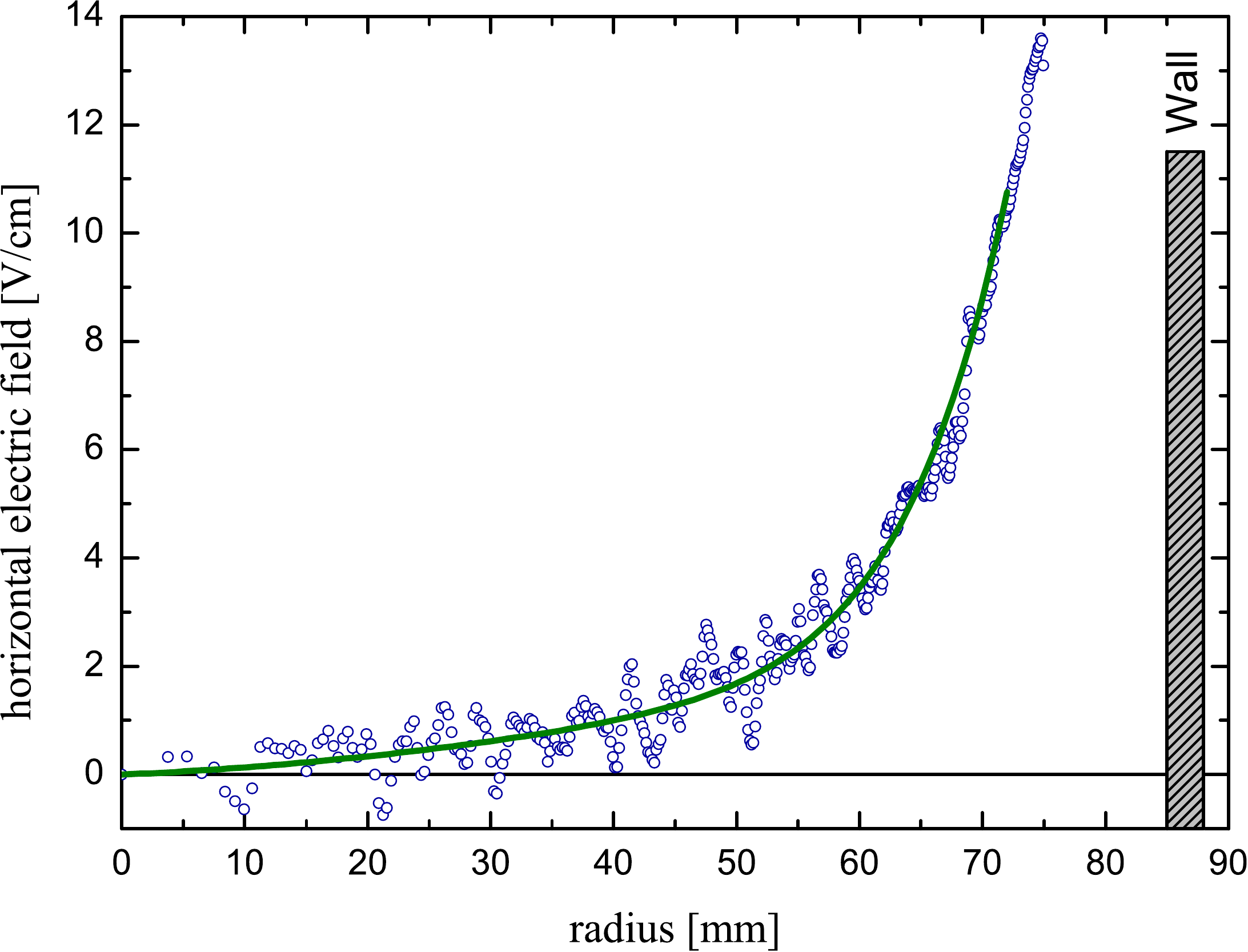}
\caption{\label{fig:comp} 
(Color online) Comparison of the radial electric field obtained from the experiment (solid line, same as in fig.~\ref{fig:9um}) and the PIC/MCC simulation (circles, same as in fig.~\ref{fig:pic}(d).} 
\end{figure}

Figure~\ref{fig:comp} shows the comparison between the measured and the computed radial electric field. Remarkable agreement can be seen. We have to emphasize that the only free linear scaling parameter is the electric charge of the dust grain, which was found to be $Z=q/e=-13500\pm500$, which is consistent with our first estimate.  

\subsection{Glass box application}

After having demonstrated the consistency of our experimental and numerical results, we apply this method to an experimental system frequently used since the discovery of Yukawa balls \cite{Cball}, which are spherical (three dimensional) ensembles of dust particles with interesting new features, like ordered shell structures \cite{YBshell} and anisotropic melting properties \cite{YBmelt}, and other vertical dust structures \cite{YBstring}. Common to these experiments is the use of an open (on top and bottom) glass box to modify the confinement field and other plasma parameters. In principle, from point of view of symmetry, the use of a cylinder instead of the box would be more appropriate, but due to practical experimental reasons a glass cylinder, which would induce unwanted distortions in the optical tracking of the grains, is clearly not preferred. For our numerical simulations, however, the glass box has to be replaced by a cylinder, otherwise a full three dimensional PIC/MCC simulation would be needed, which requires extraordinary computational resources. Luckily, most glass box experiments report principally spherically or cylindrically symmetric dust ensembles. It seems to be controversial that spherical structures are formed in a field, which is determined by cubic boundary conditions. We have conducted an experiment campaign with the aim to map the field inside a glass box more precisely to show the transition from the cylindrical to rectangular symmetry of the confinement field inside the glass box.

\begin{figure}[htb]
\centering
\includegraphics[width=0.5 \columnwidth]{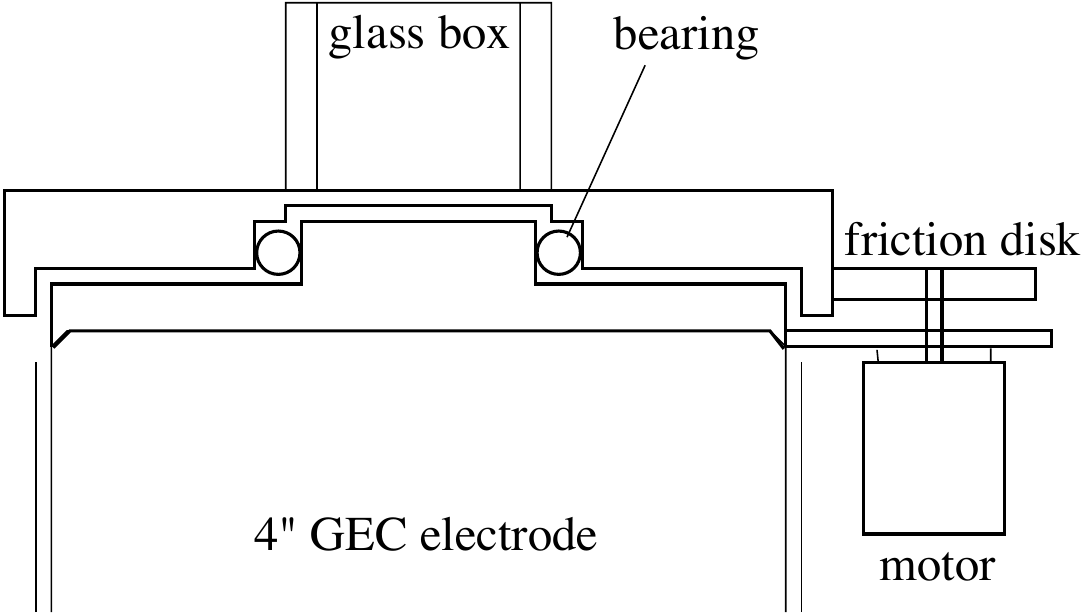}
\caption{\label{fig:rotext} 
Schematics of the rotatable extension placed on the standard 4 inch electrode column of the GEC reference cell.} 
\end{figure}

To realize this experiment we have modified the dusty plasma setup at the Hypervelocity Impacts and Dusty Plasmas Lab (HIDPL) of the Center for Astrophysics, Space Physics, and Engineering Research (CASPER) at Baylor University, Waco, Texas. Detailed description of the setup is given in ref.~\cite{Boess04}. We have added a rotatable extension to the powered lower electrode, driven by a small dc motor directly attached to it, as schematically shown in figure~\ref{fig:rotext}. A glass box with inner side length $L=18$~mm was fixed centered on top of the rotatable electrode disc. The argon discharge at a pressure of $p=20$~Pa was driven by 10 Watts of RF~power. Dust grains with diameter $a=6.5~\mu$m were inserted into the box forming a small, cylindrically symmetric single layer plasma crystal. 

\begin{figure}[htb]
\centering
\includegraphics[width=0.9 \columnwidth]{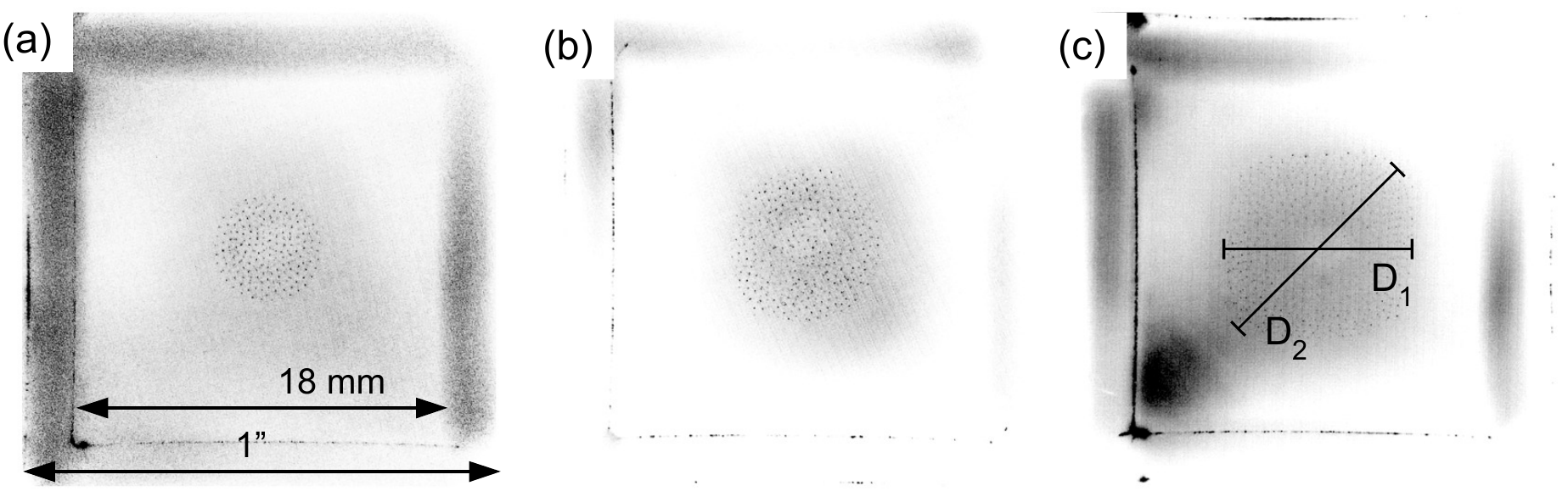}
\caption{\label{fig:exp2} 
Inverted snapshots of the rotating glass box experiments. The rotation speeds are (a) 0 rot/s, (b) 2.66 rot/s, and (c) 4.0 rot/s. In addition in (a) the dimensions are shown, while in (c) the parallel and diagonal diameters are illustrated.} 
\end{figure}

\begin{figure}[htb]
\centering
\includegraphics[width=0.7 \columnwidth]{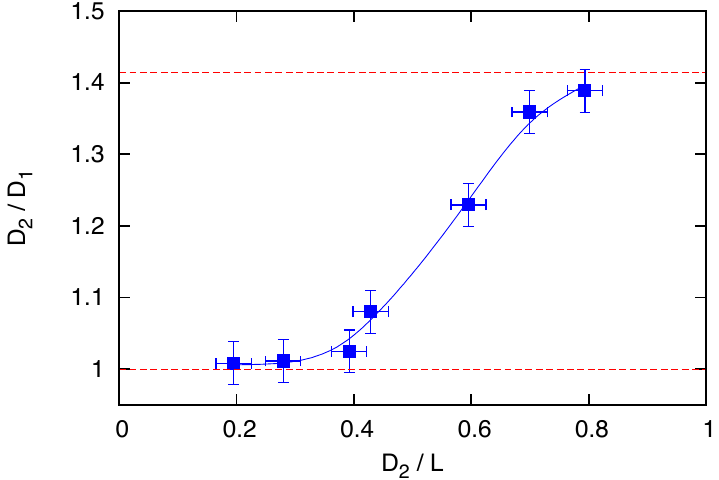}
\caption{\label{fig:diag} 
(Color online) Asymmetry parameter $D_2/D_1$ vs. the ratio of the diagonal diameter to box the side length. Lower and upper dashed lines indicate theoretical limits.} 
\end{figure}

Figure~\ref{fig:exp2} shows examples of particle snapshots at three different rotation rates. The video sequences were taken from the top at 125 frames per second (fps) with a short exposure time of 1/5000~s. At slow rotation speeds the dust cloud extends radially outward and shows a cylindrically symmetric structure, while at fast rotation, the cloud extends further towards the walls of the glass box, where the confinement field approaches a rectangular symmetry. To quantify the anisotropy, we measure the apparent diameter of the dust cloud along directions parallel to the glass box walls $(D_1)$ and its diagonal $(D_2)$. The ratio of these two diameters is plotted in fig.~\ref{fig:diag} versus the ratio $D_2/L$ ratio.

Based on these results we conclude, that (at least under the particular condition of this experiment) dust particles inside the glass box, situated within the central 40\% of the linear glass box side length, experience a cylindrically symmetric confinement field. Further out of this region the confinement field starts following the square symmetry of the glass box. The use of a glass cylinder, replacing the box, would provide symmetric confinement over the whole radial extend, while the detection could be performed with our new, single camera 3D imaging method \cite{Lytro}, which could observe the dust cloud from the top, as it is done in traditional single layer experiments.

\section{Summary}

With the RotoDust setup, shown schematically in figure~\ref{fig:setup}, we are able to continuously scan over the horizontal levitation plane of the dust grain and precisely measure the distribution of the horizontal (confinement) field experienced by the dust grains. In our first experiment we have demonstrated a sensitivity of approximately 0.1~V/cm in electric field measurements. Comparison with our cylindrically symmetric PIC/MCC simulation resulted in excellent agreement between experimental and numerical results with only a single linear scaling parameter, the dust grain charge. The RotoDust principle was used to show the size limit where cylindrically symmetric confinement fields in glass box experiments can be expected.

\ack
 This research has been supported by the OTKA Grant NN-103150.

\section*{References}
%\bibliographystyle{unsrt}
%\bibliography{rotbib}

\end{document}